\documentclass[twocolumn,showpacs,preprintnumbers,amsmath,amssymb]{revtex4}
\usepackage{dcolumn}
\usepackage{bm}
\expandafter\ifx\csname package@font\endcsname\relax\else
 \expandafter\expandafter
 \expandafter\usepackage
 \expandafter\expandafter
 \expandafter{\csname package@font\endcsname}%
\fi

\usepackage{graphicx}   
\usepackage{amsmath}    

\newcommand{\ket}[1]{\mbox{$| #1 \rangle$}}

\begin{document}


\title{Decoy State Quantum Key Distribution}

\author{Hoi-Kwong Lo, Xiongfeng Ma, and Kai Chen}

\affiliation{%
Center for Quantum Information and Quantum Control,\\
Dept. of Electrical \& Computer Engineering and Dept. of Physics,\\
University of Toronto, Toronto, Ontario, M5S 3G4, CANADA\\
}%

\date{\today}

\begin{abstract}
There has been much interest in quantum key
distribution. Experimentally, quantum key distribution over 150 km
of commercial Telecom fibers has been successfully performed. The
crucial issue in quantum key distribution is its security.
Unfortunately, all recent experiments are, in principle, insecure
due to real-life imperfections. Here, we propose a method that
can for the first time make most of those experiments secure by
using essentially the same hardware. Our method is to use decoy
states to detect eavesdropping attacks. As a consequence, we have
the best of both worlds---enjoying unconditional security guaranteed
by the fundamental laws of physics and yet dramatically surpassing
even some of the best experimental performances reported in the
literature.
\end{abstract}

\maketitle

Quantum key distribution (QKD) allows two users, Alice and Bob, to
communicate in absolute security in the presence of an eavesdropper,
Eve. Unlike conventional cryptography, the security of QKD is based
on the fundamental laws of physics, rather than unproven
computational assumptions. The security of QKD has been rigorously
proven in a number of recent papers
\cite{mayersqkd}. See also \cite{security}.
There has been tremendous
interest in experimental QKD \cite{GYS,NEC}, with the current world
record distance of 150 km of Telecom fibers\cite{NEC}.

Unfortunately, all those exciting recent experiments are, in principle,
insecure due to real-life imperfections. More concretely,
highly attenuated lasers are often used as sources. But, these
sources sometimes produce signals that contain more than one
photons. Those multi-photon signals open the door
to powerful new eavesdropping attacks including photon splitting attack.
For example, Eve can, in principle, measure the photon number of
each signal emitted by Alice and selectively suppress single-photon
signals. She splits multi-photon signals, keeping one copy for herself
and sending one copy to Bob. Now, since Eve has an identical copy
of what Bob possesses, the unconditional
security of QKD (in, for example, standard
BB84 protocol\cite{BB84}) is completely compromised.

In summary, in standard BB84 protocol, only signals originated from
{\it single} photon pulses emitted by Alice are guaranteed to be
secure. Consequently, paraphrasing GLLP \cite{GLLP}, the secure key
generation rate (per signal state emitted by Alice) can be shown to
be given by:
\begin{equation}\label{refinedkeyrate}
S \geq Q_{\mu} \{- H_2(E_{\mu}) + \Omega [1-H_2(e_1)]\}
\end{equation}
where $Q_{\mu}$ and $E_{\mu}$ are respectively the gain and quantum
bit error rate (QBER) of the signal state \cite{gain}, $\Omega$ and
$e_1$ are respectively the fraction and QBER of detection events by
Bob that have originated from single-photon signals emitted by
Alice and $H_2$ is the binary Shannon entropy.

It is a priori very hard to obtain a good lower bound on $\Omega$
and a good upper bound on $e_1$.
Therefore, prior art methods (as in GLLP \cite{GLLP}) make the
most pessimistic assumption that all multi-photon signals emitted
by Alice will be received by Bob. For this reason,
until now, it has been widely believed that the demand for
unconditional security will severely reduce the performance of QKD
systems \cite{ilm,GLLP,gisin,prior,koashi}.

In this paper, we present a simple method that will provide very good
bounds to $\Omega $ and $e_1$. Consequently, our method for the first time makes
most of the long distance QKD experiments reported in the literature
unconditionally secure. Our method has the advantage that it can be
implemented with essentially the current hardware. So, unlike prior
art solutions based on single-photon sources, our method
does not require daunting experimental
developments. Our method is based on the decoy state idea first
proposed by Hwang \cite{hwang}. While the idea of Hwang
was highly innovative,
his security analysis was heuristic.
The key point of the decoy state
idea is that Alice prepares a set of additional states---decoy
states, in addition to standard BB84 states. Those decoy states are
used for the purpose of detecting eavesdropping attacks only,
whereas the standard BB84 states are used for key generation only.
The only difference between the decoy state and the standard BB84
states is their intensities (i.e., their photon number
distributions).

By measuring the yields and QBER of
decoy states, we will show that Alice and Bob can obtain
reliable bounds to $\Omega$ and $e_1$, thus
allowing them to surpass all prior art results substantially
\cite{entanglement}. Here,
we give for the first time a rigorous analysis of the
security of decoy state QKD.
Moreover, we show that the decoy state idea can be combined with
the prior art GLLP \cite{GLLP} analysis.

Preliminary versions of our result in this paper have appeared in
\cite{preliminary1,preliminary2}, where
we presented not only the general theory, but also proposed the idea
of using only a few decoy states (for example,
three states---the vacuum, a weak decoy state with $\mu_{decoy} \ll 1$
and a signal state with $\mu = O (1)$. We call this
a Vacuum+Weak decoy state protocol).
Subsequently, our protocols
for decoy state QKD have been analyzed in
\cite{wangdecoy} and more systematically
in \cite{practicaldecoy}. See also \cite{harrington}.
Recently, we have
provided the first experimental demonstration of decoy state QKD in
\cite{experimentaldecoy}.

We now present the general theory of our new
decoy state schemes.
We will assume that Alice can prepare phase-randomized
coherent states and can turn her power up and down for
each signal. This may be achieved
by using standard commercial variable optical attenuators (VOAs) \cite{voas}.
Let $\ket{\sqrt{\mu} e^{i \theta}}$ denote a weak coherent state
emitted by Alice. Assuming that the phase, $\theta$, of all signals
is totally randomized,
the probability distribution for the
number of photons of the signal state follows a Poisson distribution
with some parameter $\mu$.
That is to say that, with a probability $p_n = e^{ - \mu} \mu^n / n!$,
Alice's signal will have $n$ photons.
In summary, we have assumed that Alice can prepare any Poissonian (with
parameter $\mu$)
mixture of photon number states and, moreover, Alice can
vary the parameter, $\mu$, for each individual signal.

Let us consider the gain $Q_{\mu}$ for a coherent
state $\ket{\sqrt{\mu} e^{i \theta}}$.
[Here and thereafter, we actually mean the
random mixture of $\ket{\sqrt{\mu} e^{i \theta}}$
over all values of $\theta$ as the phase is assumed to be
totally randomized.]
We have:
\begin{eqnarray} \label{Decoy:QSingal}
Q_{\mu} &=&  Y_0 e^{- \mu}  + Y_1 e^{- \mu} \mu   + Y_2 e^{- \mu}
\left(\mu^2/2 \right)
\nonumber \\
           &~&+\ldots+Y_n e^{- \mu} \left( \mu^n/n! \right)  + \ldots
\end{eqnarray}
where $Y_n$ is the yield of an $n$-photon signal \cite{yn} and where
$Y_0 \geq 0$ gives the
detection events due to background including dark counts
and stray light from timing pulses.

Similarly, the QBER can depend on the photon number.
Let us define $e_n$ as the QBER of an n-photon signal.
The QBER $E_{\mu}$ for a coherent state $\ket{\sqrt{\mu} e^{i \theta}}$
is given by
\begin{eqnarray}
Q_{\mu} E_{\mu} &=&  Y_0 e^{- \mu} e_0  + Y_1 e^{- \mu} \mu e_1  +
Y_2 e^{- \mu} \left(\mu^2/2 \right)  e_2
\nonumber \\
           &~&+\ldots+Y_n e^{- \mu} \left( \mu^n/n! \right) e_n  + \ldots ,
\label{signalQBER}
\end{eqnarray}
which is the weighted average of the QBERs of various photon number eigenstates.

\paragraph*{Essence of the decoy state idea}
Let us imagine that a decoy state and a signal state
have the same characteristics (wavelength, timing information, etc).
Therefore,
Eve cannot distinguish a decoy state from a signal state
and the only piece of information available to Eve is the number of
photons in a signal.
Therefore, the yield, $Y_n$, and QBER, $e_n$, can
depend on only the photon number, $n$, but not
which distribution (decoy or signal) the state is from.
We emphasize that the essence of the decoy state idea can be summarized
by the following two equations:
\begin{eqnarray}
Y_n (signal) &= &Y_n (decoy)= Y_n \\
e_n (signal) &=& e_n (decoy) = e_n .
\label{sameerror}
\end{eqnarray}

While a few decoy states are sufficient, for ease of discussion, we
will for the moment consider the case where
Alice will pick an infinite number of possible intensities
for decoy states. Let us imagine that Alice
varies over all
non-negative values of $\mu$ randomly and independently
for each signal, Alice and Bob can experimentally
measure the yield $Q_{\mu}$ and the QBER $E_{\mu}$.
Since the relations between the variables $Q_{\mu}$'s
and $Y_n$'s and between
$E_{\mu}$'s and $e_n$'s are linear, given the set of variables
$Q_{\mu}$'s and $E_{\mu}$'s
measured from their experiments,
Alice and Bob can deduce mathematically
with high confidence the variables $Y_n$'s and $e_n$'s.
This means that Alice and Bob can constrain simultaneously the yields,
$Y_n$
and QBER $e_n$ simultaneously for {\it all} $n$.
Suppose Alice and Bob know their channel property well.
Then, they know what range of values of $Y_n$'s and $e_n$'s is acceptable.
Any attack by Eve that will change the value of any one of the
$Y_n$'s and $e_n$'s
substantially will, in principle, be caught with high probability
by our decoy state method. Therefore, in order to avoid being
detected, the eavesdropper, Eve, has very limited options in her
eavesdropping attack.
In summary, the ability for Alice and Bob to verify
experimentally the values of
$Y_n$ and $e_n$'s in the decoy state method greatly strengthens
their power in detecting eavesdropping, thus leading to
a dramatic improvement in the performance of their QKD system.

The decoy state method allows Alice and Bob to detect
deviations from the normal behavior due to eavesdropping attacks.
Therefore, in what follows, we will consider normal behavior (i.e.,
the case of no eavesdropping). Details of QKD set-up
model can be seen in \cite{practicaldecoy}.

\subparagraph*{Yield}

Let us discuss the yields, $Y_n$'s, in a realistic set-up.

(a) The case $n=0$.

In the absence of eavesdropping, $Y_0$ is simply given by the
background detection event rate $p_{dark}$ of the system.

(b) The case $n \geq 1$. For $n \geq 1$,  yield $Y_n$ comes
from two sources, i) the detection of signal photons $\eta_n$,
and ii) the background event $p_{dark}$.
The combination gives, assuming the independence of background
and signal detection events,
\begin{eqnarray}\label{yieldn}
Y_n &=& \eta_n+p_{dark}-\eta_n \cdot p_{dark} \nonumber \\
&\approx & \eta_n+p_{dark}
\end{eqnarray}
where in the second line we neglect the cross term because the
background rate (typically $10^{-5}$) and transmission efficiency
(typically $10^{-3}$) are both very small.

Suppose the overall transmission probability of each photon is
$\eta$. In a normal channel, it is
common to assume independence between the behaviors of the \emph{n
photons}. Therefore, the transmission efficiency for n-photon
signals $\eta_n$ is given by:
\begin{equation}
\eta_n = 1 - ( 1 - \eta)^n ,
\end{equation}
[For a small $\eta$ and ignore the dark count,
$Y_n \approx n \eta$.]

\subparagraph*{QBER}
Let us discuss the QBERs,
$e_n$'s, in a realistic experiment.

(a) If the signal is a vacuum, Bob's detection is due to
background including dark counts and stray light due to
timing pulses. Assuming that the two detectors have
equal background event rates,
then the output is totally random
and the error rate is 50\%.
That is, the QBER for the vacuum $e_0 = 1/2$.

(b) If the signal has $n \geq 1$ photons,
it also has some error rate, say $e_n$.

More concretely, $e_n$ comes from two parts, erroneous detections
and background contribution,
\begin{equation}\label{e_n}
e_n=(e_{detector}\cdot\eta_n+\frac12p_{dark})/Y_n ,
\end{equation}
where $e_{detector}$ is independent of $n$.

The values of $Y_n$ and $e_n$ can be experimentally verified by
Alice and Bob using our decoy state method. Any attempt by Eve to
change them significantly will almost always be caught.

\paragraph*{Combining decoy state idea with GLLP}

Suppose key generation is done on signal state $\ket{\sqrt{\mu} e^{i
\theta}}$. In principle, Alice and Bob can isolate the single-photon
signals and apply privacy amplification to them only. Therefore,
generalizing the work in GLLP, we find Eq. \eqref{refinedkeyrate}
where the gain of the signal state, $Q_{\mu}= \sum_{k=0}^{\infty}
Y_k e^{- \mu} \left( \mu^k /k! \right) ,  $ [This comes directly
from Eq. \eqref{Decoy:QSingal}.] and the fraction of Bob's detection
events that have originated from single-photon signals emitted by
Alice is given by:
\begin{equation}
\Omega = { Q_1  \over
      Q_{\mu} } ,
\end{equation}
where
\begin{equation}
Q_1=Y_1\cdot\mu e^{-\mu}
\end{equation}
is the gain for the single photon state.

The derivation of Eq. \eqref{refinedkeyrate} assumes that error
correction protocols can achieve the fundamental (Shannon) limit.
However, practical error correction protocols are generally
inefficient. As noted in \cite{cascade}, a simple way to take this
inefficiency into account is to introduce a function, $f(e)> 1$, of
the QBER, $e$. By doing so, we find that the key generation rate for
practical protocols is given by:
\begin{equation} \label{practicalkeyrate}
     S \geq q  \{-Q_{\mu}f(E_{\mu})H_2(E_{\mu})+Q_1[1-H_2(e_1)]\},
\end{equation}
where $q$ depends on the implementation (1/2 for the BB84 protocol,
because half the time Alice and Bob bases are not compatible, and if
we use the efficient BB84 protocol \cite{eff}, we can have
$q\approx1$. For simplicity, we will take $q=1$ in this paper.),
and $f(e)$ is the error correction efficiency
\cite{cascade}.

Let us now compare our result in Eq. \eqref{practicalkeyrate} with
the prior art GLLP result. In the prior art GLLP \cite{GLLP} method,
secure key generation rate is shown to be at least
\begin{equation}\label{GLLPkeyrate}
S \ge Q_{\mu} \left\{ - H_2 (E_{\mu}) + \Omega \left[ 1 -
H_2(\frac{E_{\mu}}{\Omega}) \right]  \right\},
\end{equation}
where $\Omega$, the fraction of ``untagged'' photons, (which is a
pessimistic estimation of the fraction of detection events by Bob
that have originated from single-photon signals emitted by Alice),
is given by
\begin{equation}
1- \Omega = p_{multi} / Q_{\mu} , \label{e:worstcase}
\end{equation}
where $p_{multi}$ is the probability of Alice's emitting a
multi-photon signal. Eq. \eqref{e:worstcase} represents the worst
situation where all the multi-photons emitted by Alice will be
received by Bob.

Comparing our result (given in Eq. \eqref{practicalkeyrate}) with
the prior art GLLP result (given in Eq. \eqref{GLLPkeyrate}), we see
that the main difference is that in our result, a much better
lower bound on $\Omega$ and
a much better upper bound on $e_1$ can be obtained.

\begin{figure}
\centering \resizebox{8cm}{!}{\includegraphics{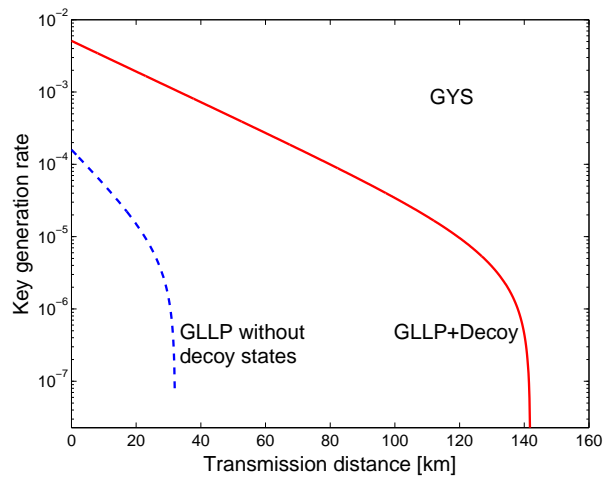}}
\caption{This figure is obtained by writing a simple program. The
line of GLLP without decoy state uses the formula
(\ref{GLLPkeyrate}) and GLLP+Decoy uses the formula
(\ref{practicalkeyrate}) according to the parameters given in
experiment GYS \cite{GYS}. With decoy states, the maximal distance
increases to over 140 km. For comparison, we found that with prior
art method, the secure distance is only about 30 km. We have also
proven that an upper bound of distance of secure BB84 with the GYS
parameters is 208 km because this corresponds to the point where
$e_1 = 1/4$ and the protocol is insecure due to an intercept-resend
attack. We have checked that our results are stable to small
perturbations to the background event rate $p_{dark}$ and average
photon number $\mu$ (both up to 20\% change). } \label{fig1}
\end{figure}

\paragraph*{Implication of our result}

We obtain substantially higher key generation rate than in
\cite{GLLP}. In more detail, note that, from Eq. \eqref{yieldn},
$Y_n$ for $n>2$ is of similar order to $Y_1$. Therefore, from Eq.
\eqref{practicalkeyrate} it is now advantageous for Alice to pick
the average photon number in her signal state to be $\mu = O( 1)$.
Therefore, the key generation rate in our new method is $ O (\eta)$
where $\eta$ is the overall transmission probability of the channel.
In comparison, in prior art methods for secure QKD, $\mu$ is chosen
to be of order $O (\eta)$, thus giving a net key generation rate of
$O (\eta^2)$. In summary, we have achieved a substantial increase in
net key generation rate from $O (\eta^2)$ to $O (\eta)$. Moreover,
as will be discussed below, our decoy state method allows secure QKD
at much longer distances than previously thought possible.

More concretely, we \cite{preliminary2} have applied our results to various experiments
in the literature.  The results are shown in
Fig.~1 using the GYS \cite{GYS} experiment as an example. We found
that the optimal averaged number $\mu$ in GYS that maximizes the key
generation rate in our decoy state method in Eq.
\eqref{practicalkeyrate} is, indeed, of $O(1)$ (roughly 0.5).
Therefore, the key generation rate is of order $O(\eta)$. We remark
that the calculated optimal value of photon number of 0.5 is, in
fact, {\it higher} than what experimentalists have been using.
Experimentalists often liberally pick 0.1 as a convenient number for
average photon number without any security justification. In other
words, operating their equipment with the parameters proposed in the
present paper will allow experimentalists to not only match, but
also {\it surpass} their current experimental performance (by having
at least five-fold the current experimental key generation rate).
This demonstrates clearly the power of decoy state QKD. Moreover,
Fig.~1 shows that with our decoy state idea, secure QKD can be done
at distances over 140 km with only current technology.

In summary, our result shows that we can have the best of
both worlds: Enjoy both unconditional security and record-breaking
experimental performance.
The general principle of decoy state
QKD developed here can have widespread applications in other
set-ups (e.g. open-air QKD or QKD with other photon sources)
and to multi-party
quantum cryptographic protocols such as \cite{ChenLo}.
As demonstrated
clearly in \cite{practicaldecoy},
one can achieve
almost all the benefits of our decoy state method with only one or
two decoy states. See also \cite{wangdecoy}.
Recently, we have experimentally demonstrated decoy
state QKD in \cite{experimentaldecoy}.

We have benefitted greatly from enlightening discussions with
many colleagues including particularly
G. Brassard.
Financial support from funding agencies such as CFI,
CIPI, CRC program, NSERC, OIT, and PREA are gratefully acknowledged.
H.-K. L also thanks travel support from the INI,
Cambridge, UK and from the
IQI at Caltech
through NSF grant EIA-0086038.



\begin{thebibliography}{99}
\bibitem{gain} Here, the gain means the ratio of the number of
Bob's detection events (where Bob chooses
the same basis as Alice) to Alice's number of emitted signals.
QBER means the error rate of Bob's
detection events for the case that Alice and Bob use the
same basis.
\bibitem{voas} Variable Optical Attenuators (VOAs) at repetition rates as high
as 100GHz are commercially widely available. Those attenuators can be directly
applied to fiber-optics based QKD systems.
\bibitem{yn} $Y_n$ is
the conditional probability that Bob detects a signal, given
that an $n$-photon signal is emitted by Alice.
\end{thebibliography}
\end{document}